\documentclass[aps,prl,showpacs,twocolumn]{revtex4}
\usepackage[british]{babel}
\usepackage{amsmath, amssymb}
\usepackage{graphicx}

\begin{document}
\title{Dirac's Quantum Phase Problem} 
\author{J. Sperling} \email{jan.sperling2@uni-rostock.de}
\author{W. Vogel} \email{werner.vogel@uni-rostock.de}
\affiliation{Arbeitsgruppe Quantenoptik, Institut f\"ur Physik, Universit\"at Rostock, D-18051 Rostock, Germany}
\date{\today}

\begin{abstract}
	In 1927 the great physicist Paul A. M. Dirac failed to provide a consistent quantum description of the phase of a radiation field.
	Only one year later, he developed the famous Dirac theory of the electron, which led to the anti-particle -- the positron.
	We show that the reason for Dirac's failure with the phase problem bears a striking resemblance to his ingenious insight into the nature of the electron.
	For a correct quantum description of the phase of a radiation field it is necessary to take the polarisation into account.
	Similarly to the introduction of the anti-particle of the electron, the inclusion of the second polarisation resolves the inconsistency of the quantum phase problem. This also leads to new insight into the quantum measurement problem of time.
\end{abstract}

\maketitle

The transition from the classical to the quantum description of radiation fields and matter systems became of increasing importance during the last decades.
This development was stimulated by the improvements of experimental techniques, in particular the rapidly improving laser
sources.
The coherence properties of the lasers also led to new developments of phase-sensitive measurement techniques.
From the fundamental point of view the quantum description of the phase properties of a radiation field was of interest already some decades before the experimental techniques allowed their observation.

In the framework of quantum field theory the complex amplitudes $\alpha$ and $\alpha^\ast$  of a radiation mode are replaced by the photon annihilation and creation operators $\hat a$ and $\hat a ^\dagger$, respectively.
The correspondence between the classical complex amplitude, $\alpha=|\alpha|\exp({\rm i}\phi)  $, and the non-Hermitian operator $\hat a$ led to the question for a proper description of the quantum properties of the phase.
Dirac defined a Hermitian phase operator $\hat \Phi$ by translation of the classical relation to the   quantum domain~\cite{prsla114-243}:
\begin{align}
	\hat a=  \exp(i\hat \Phi) \, \hat n ^{1/2}, \intertext{with the commutator} \quad [\hat \Phi,\hat n]=-{\rm i},
\end{align}
where $\hat n=\hat a^\dagger\hat a$ is the photon number operator.
Based on these assumptions it is easily seen that the commutation rule leads to ill-defined diagonal matrix elements of the phase operator in the photon-number basis, $\langle n | \hat \Phi |m \rangle = - {\rm i} \delta_{n,m} /(m-n)$.
This and some related problems with the proper definition of an observable (and thus Hermitian) phase operator led to a long lasting debate.
Here we may only briefly discuss some of its main results.

An exponential phase operator, $\hat V_{+} = \widehat{e^{{\rm i} \Phi}}= \sum_{n=0}^\infty |n\rangle \langle n+1|$, was defined by Susskind and Glogower~\cite{physics1-49}. This operator is no longer unitary, nor it is directly expressed in terms of a phase operator $\hat \Phi$.
It is easily seen that the commutator, $[\hat V_{+}, \hat V_{+}^\dagger] = |0\rangle \langle 0|$, leads to clear signatures of the non-unitary operator whenever the quantum state of the system has a non-vanishing projection onto the photon vacuum.
The solution of the eigenvalue problem of $\hat V_{+}$, $\hat V_{+} |\phi \rangle = e^{{\rm i}\phi}|\phi \rangle$, leads to the phase states $|\phi \rangle_{+} = (2\pi)^{-1/2} \sum_{n=0}^\infty e^{{\rm i}n\phi} |n\rangle$ as introduced by London~\cite{zphys37-915}.
Based on these states one may calculate the London phase distributions that show some signatures of quantum phase distributions, but they do not represent a properly defind probability of a phase observable.
Nevertheless, these phase distributions are experimentally accessible~\cite{oc148-355}, by using methods of quantum state reconstruction, for a review see~\cite{proopt39,rmp81-299}.

Other approaches to the quantum description of the phase are based on Hermitian cosine and sine operators, for details see~\cite{rmp40-411}.
The properties of the quantum phase obtained in this manner do not correspond to the existence of an unambiguously defined phase operator.
Two different proposals of introducing Hermitian phase operators have been given by Barnett and Pegg.
In the first one they artificially extend the $n$-summation in the exponential phase operator $\hat V$ to $-\infty$, cf.~\cite{jphysa19-3849}.
This extension was justified by the argument that no physical interaction would couple between the positive and negative photon-number states.
In their second approach~\cite{epl6-483}, Pegg and Barnett choose discrete phase values by truncating the Fock basis in the representation of the London phase state $|\phi \rangle_{+}$.
The resulting phase distributions, which are attributed by the authors to a formally Hermitian phase operator, are equal to the distributions resulting from the London phase states, which do not represent proper phase distributions.
In fact, the used truncation procedure is valid only for bounded operators, but not for the unbounded ones for which it was applied.
For a detailed review on the quantum phase problem we also refer to~\cite{pr256-367}.

Now we are going to introduce a correct version of a Hermitian phase operator that is eventually consistent with Dirac's conjecture.
For simplicity, let us consider a monochromatic electromagnetic plane-wave field freely propagating in the positive $x$ direction.
Its transverse electric field vector is described in the Heisenberg picture, up to a constant factor, by the operator ${\bf \hat E} (x,t) \propto i\left({\bf \hat a} e^{i(kx-\omega t +\phi_0)}- {\bf \hat a}^\dagger e^{-i(kx-\omega t+\phi_0)}\right )$, with
\begin{align*}\nonumber
 	{\bf \hat a}&=\sum_{\mu=\pm}  {\bf e}_\mu \hat a_\mu=
	\left(\begin{array}{c}
		\hat a_{+}\\ \hat a_{-}
	\end{array}\right)\\
	&=\hat a_{+}\otimes|\sigma_+\rangle\langle\sigma_+|+\hat a_{-}\otimes|\sigma_-\rangle\langle\sigma_-|,
\end{align*}
where ${\bf e}_\mu$ is the polarisation unit vector with $\mu =\pm$ for the $\sigma_\pm$ circular polarisations.
For convenience let us call $\sigma_{+}$ the circular polarisation, and $\sigma_{-}$ the anti-circular polarisation, cf. Fig.~1.
Further on, for any operator depending on the polarisation we will use the boldface vector notation.
The constant $\phi_0$ defines the initial phase at position $x=0$ of the mode structure, $\omega$ is the frequency of the field.
The term $\hat a_{+}\otimes|\sigma_+\rangle\langle\sigma_+|$ stands for the circularly polarised frequency mode, and $\hat a_{-}\otimes|\sigma_-\rangle\langle\sigma_-|$ for the anti-circularly polarised one.

\begin{center}
\begin{figure}
	\includegraphics[scale=0.8]{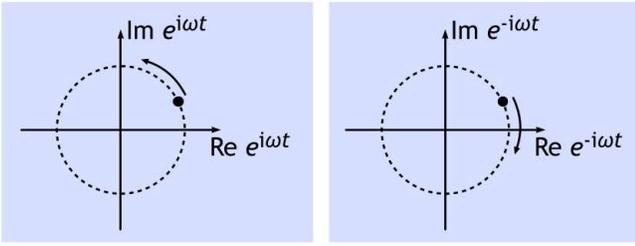}
	\caption{
		The evolution of the circular and anti-circular polarisations is outlined, which are related to one another by inverting the direction of time, or by changing the sign of the frequency.
	}
\end{figure}
\end{center}

The usual approaches for introducing phase operators proceed to consider a single mode of the field. At this point we arrive at the observation that the space-time dependent phase of the wave, $kx -\omega t +\phi_0$, is obviously related to both polarisations of the field vector.
Thus we conclude that the phase properties are related to both polarisation components of the monochromatic field.
Hence the deep problems of the quantum description of the phase may result from the impermissible reduction of the problem to a single mode field.

The Hamiltonian for a single radiation mode -- say the circularly polarised $\sigma_+$ mode -- reads as $\hat H_{+}=
\hbar \omega\sum_{n_+\in\mathbb{N}} \left(n_++\frac{1}{2}\right)|n_+\rangle\langle n_+|$.
The polarisation ($|\sigma_{+}\rangle$ for circular, and $|\sigma_{-}\rangle$ for anti-circular) represents an additional degree of freedom.
Thus the compound Hilbert space for a monochromatic field of  frequency $\omega$ and the polarisation is given by $\mathcal H=\mathcal H_{+}\otimes\mathbb C^2$.
In fact, formally we may assign to the circularly polarised photon the anti-circularly polarised photon of the same frequency as its anti-particle.
The ''anti-photon'' can be alternatively related to a negative amount of energy, or a hole in the corresponding ''Dirac sea''.
An isomorphism without changing the ordering -- with respect to the energy -- is the following:
\begin{align}\nonumber
	|n,\sigma\rangle\mapsto\left\lbrace
	\begin{array}{lcc}
		|\mathcal{E}_n\rangle & \mbox{for} & \sigma=\sigma_{+},\\
		|\mathcal{E}_{-n-1}\rangle & \mbox{for} & \sigma=\sigma_{-}.
	\end{array}
	\right.
\end{align}
For the new Hilbert space we obtain, in the basis $\{|\mathcal{E}_n\rangle:n\in\mathbb{Z}\}$, the Hamiltonian in the form $\hat {\bf H}=\sum_{n\in\mathbb{Z}} \mathcal{E}_n|\mathcal{E}_n\rangle\langle \mathcal{E}_n|,$ together with $\mathcal{E}_n=\hbar \omega\left(n+\frac{1}{2}\right)$ and the number operator
\begin{align}
	\hat {\bf n}=\sum_{n\in\mathbb{Z}} n|\mathcal{E}_n\rangle\langle \mathcal{E}_n|
\end{align}
For the anti-circularly polarised photons we obtain the desired negative energy amount. The physical interpretation is straightforward: adding a circularly polarised photon to a monochromatic field is equivalent to subtracting an anti-circularly polarised one. It also becomes clear that for our purposes circular polarisation is more fundamental than linear one.
In fact, a vertically polarised photon cannot be formally considered to be the antiparticle of the horizontally polarised one.

Already here it becomes obvious that our approach to the quantum phase problem bears a striking resemblance to the Dirac theory of the relativistic electron~\cite{dirac}.
The structure of the Dirac equation led to a consistent quantum theory of the relativistic electron which includes both the spin of the electron and its anti-particle -- the positron.
The quantum field theory of electromagnetism obeys relativistic invariance per se, here the photon polarisation (or spin) is included and the anti-particle is the photon itself.
Since the phase of the wave is related to the frequency of the field -- with a polarisation degeneracy -- it will be shown in the following that all inconsistencies of the quantum theory of the phase disappear by a consistent description of a monochromatic field as a two-mode system.

From functional analysis it is known that all Hilbert spaces with a countable basis are isomorphic~\cite{fkt-an-book}.
Thus we can choose a special one, namely the square integrable $2\pi$-periodic functions, to represent our system given in Fock basis and polarisation.
The bases of the spaces maps as $|\mathcal{E}_n\rangle \mapsto u_n(\phi)=\exp(-{\rm i}n\phi)$,
and the inner product is given by $\langle u|v\rangle=\frac{1}{2\pi}\int\limits_{-\pi}^{\pi} u^\ast(\phi)v(\phi)\,d\phi$.
We obtain the representation of the number operator in the form $\hat {\bf n}={\rm i}\frac{\partial}{\partial \phi}$,
with the desired eigenvalue problem $\hat {\bf n} u_n=nu_n$.
Obviously, the conjugated variable is $\hat{\bf \Phi}=\phi$, since for all $2\pi$-periodic functions $u(\phi)$ we obtain $[\phi,{\rm i}\frac{\partial}{\partial \phi}]u(\phi)=-{\rm i}u(\phi)$.

Using this representation, we easily obtain the correct exponential phase operator $\hat {\bf V}$ as
\begin{align}
	\nonumber\hat {\bf V}&=\exp({\rm i}\hat{\bf \Phi})=\sum_{m,n\in\mathbb Z}\langle \mathcal{E}_m|\exp({\rm i}\hat {\bf \Phi})|\mathcal{E}_n\rangle|\mathcal{E}_m\rangle\langle \mathcal{E}_n|\\&=\sum_{m\in\mathbb Z} |\mathcal{E}_m\rangle\langle \mathcal{E}_{m+1}|,
\end{align}
with $\langle \mathcal{E}_m|\exp({\rm i}\hat {\bf \Phi})|\mathcal{E}_n\rangle=\frac{1}{2\pi}\int\limits_{-\pi}^{\pi} e^{{\rm i}(m-n+1)\phi}\,d\phi$.
Now $\hat {\bf V}$ is a truly unitary operator, $\hat {\bf V}^{-1}=\hat {\bf V}^\dagger$.
Let us reconsider this operator from the viewpoint of the usual single-mode treatment. Using the  definition $\hat V_{+}=\sum_{k=1}^\infty|k-1\rangle\langle k|$, we obtain
\begin{align}
	\hat {\bf V}=\hat V_{+}\otimes|\sigma_{+}\rangle\langle \sigma_{+}|+|0\rangle\langle 0|\otimes|\sigma_{-}\rangle\langle \sigma_{+}|+\hat V_{+}^\dagger\otimes|\sigma_{-}\rangle\langle\sigma_{-}|.
\nonumber
\end{align}
This operator fulfils Dirac's commutation relation $[\hat {\bf V},\hat {\bf n}]=\hat {\bf V}$ for the photon number state and the exponential phase operator.
The eigenvalue problem of $\hat {\bf V}$ delivers the phase states
\begin{align} 
|\phi\rangle=\frac{1}{\sqrt{2\pi}}\sum_{k\in\mathbb Z}\exp({\rm i}\phi n)|\mathcal{E}_n\rangle
\end{align}
for the eigenvalue $\exp({\rm i}\phi)$.
Since $|\phi\rangle$ cannot be normalised, these orthogonal vectors must be read in terms of projection-valued measures as $\hat {\bf V}=\int\limits_{-\pi}^{\pi}\exp({\rm i}\phi)\, \left(d\phi|\phi\rangle\langle\phi|\right)$, and with $\hat {\bf 1}=\int\limits_{-\pi}^{\pi}\, d\phi|\phi\rangle\langle\phi|=\sum_{n\in\mathbb Z}|\mathcal{E}_n\rangle\langle \mathcal{E}_n|$ being the unit operator.

Based on these results we readily obtain a consistent definition of a Hermitian phase operator. By use of $\hat{\bf \Phi}=-{\rm i}\log \hat {\bf V}$ we obtain $\hat{\bf \Phi}|\phi\rangle=\phi|\phi\rangle$, and
\begin{align}
	\hat {\bf \Phi}=\int\limits_{-\pi}^{\pi} \phi \, d\phi|\phi\rangle\langle\phi|.
\end{align}
The matrix elements in the Fock basis are given by
\begin{align}
	\nonumber\langle \mathcal{E}_m|\hat{\bf \Phi}|\mathcal{E}_n\rangle&=\frac{1}{2\pi}\int\limits_{-\pi}^{\pi}\phi\exp({\rm i} (m-n)\phi)\,d\phi\\&=\left\lbrace\begin{array}{lcc}
		0 &\mbox{for}&m=n,\\
		\frac{\cos[\pi(m-n)]}{{\rm i}(n-m)}&\mbox{for}&m\neq n.
	\end{array}\right.\nonumber
\end{align}
Now the diagonal matrix elements are no longer ill-defined as in the case of a single-mode theory. 

In the following we consider the phase statistics of polarised single-mode radiation.
The probability density $p_{\hat {\bf \rho}}(\phi)$ for a quantum state $\hat \rho$ is given by $p_{\hat {\bf \rho}}(\phi)=\langle\phi|\hat\rho|\phi\rangle$.
As an example, let us restrict our consideration to states which can be decomposed as ${\hat {\bf \rho}}=\hat\rho_{+}\otimes\hat P$, with the polarisation $\hat P$.
Therefore we decompose
	$|\phi\rangle=|\phi\rangle_+\otimes|\sigma_{+}\rangle+e^{-{\rm i}\phi}|-\phi\rangle_+\otimes|\sigma_{-}\rangle$,
with $|\phi\rangle_+=\frac{1}{\sqrt{2\pi}}\sum_{k=0}^\infty\exp({\rm i}n\phi)|n\rangle$.
In this case we obtain
\begin{align}
	\nonumber p_{\hat {\bf \rho}}(\phi)=&	\underbrace{{}_{+}\langle\phi|\hat\rho_{+}|\phi\rangle_{+}\langle\sigma_{+}|\hat P|\sigma_{+}\rangle}_{\mbox{\small circular polarisation}}\\ \nonumber&+
	\underbrace{{}_{+}\langle-\phi|\hat\rho_{+}|-\phi\rangle_{+}\langle\sigma_{-}|\hat P|\sigma_{-}\rangle}_{\mbox{\small anti-circular polarisation}}\\
&+ \underbrace{2{\rm Re}\left(e^{-{\bf i\phi}}{}_{+}\langle\phi|\hat\rho_{+}|-\phi\rangle_{+}\langle\sigma_{+}|\hat P|\sigma_{-}\rangle\right)}_{\mbox{\small polarisation interference}}.
\label{eq:probability}
\end{align}
In the case of circularly polarised radiation, $\hat P=|\sigma_{+}\rangle\langle\sigma_{+}|$, we obtain $p_{\hat {\bf \rho}}(\phi)={}_{+}\langle\phi|\hat\rho_{+}|\phi\rangle_{+}=p_{+}(\phi)$, representing the London phase distribution.
For anti-circular polarisation, $\hat P=|\sigma_{-}\rangle\langle\sigma_{-}|$, we obtain $p_{\hat {\bf \rho}}(\phi)=p_{+}(-\phi)$.
Unpolarised radiation, $\hat P=\frac{1}{2}\hat 1$, delivers $p_{\hat {\bf \rho}}(\phi)=\frac{1}{2}p_{+}(\phi)+\frac{1}{2}p_{+}(-\phi)$.
If one is not interested in the polarisation effects, for any quantum state the phase distribution (obtained by tracing over polarisation) is equal to the distribution of the corresponding unpolarised state.
For horizontal or vertical polarisation, $\frac{1}{\sqrt{2}}(|\sigma_+\rangle\pm|\sigma_-\rangle)$, we obtain $p_{\hat {\bf \rho}}(\phi)=\frac{1}{2}p_{+}(\phi)+\frac{1}{2}p_{+}(-\phi)\pm{\rm Re}\left[e^{{\rm i}\phi}{}_+\langle\phi|\hat \rho_{+}|-\phi\rangle_{+}\right]$.

\begin{center}
\begin{figure}
	\includegraphics[scale=0.5]{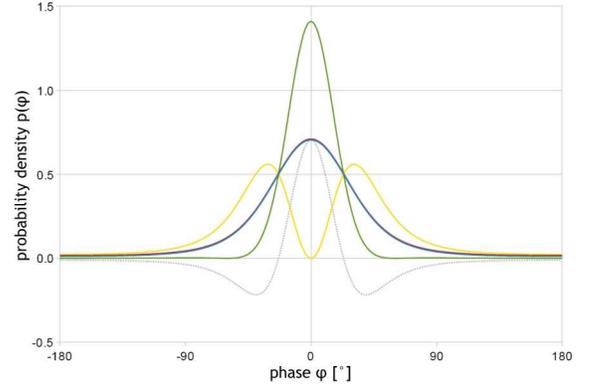}
	\caption{
		The equal red and blue lines represent a coherent state ($\alpha=1$) with circular and anti-circular polarisation, respectively.
		They are equivalent to the London phase distribution.
		The grey dotted line represents the polarisation interference term in Eq.~(\ref{eq:probability}).
		The addition / substraction of this interference term delivers the horizontal (green line) / vertical (yellow line) polarised coherent state.
	}
\end{figure}
\end{center}

\begin{center}
\begin{figure}
	\includegraphics[scale=0.5]{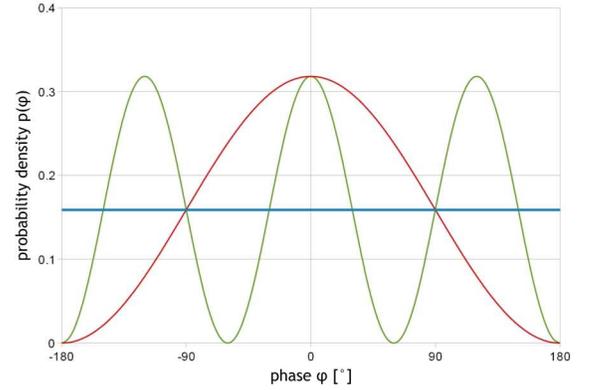}
	\caption{
		The blue line is an arbitrary circularly polarised Fock state, which has an equally distributed phase.
		However, due to the fact that the phase distribution depends on the polarisation, we obtain interferences for horizontally polarised photons, as shown for the vacuum state (red line) and the single-photon state (green line).
	}
\end{figure}
\end{center}

In Fig.~2 we consider the phase distributions for a coherent state of amplitude $\alpha=1$, which are clearly seen to depend sensitively on the polarisation of the radiation.
The phase distribution of any circularly polarised Fock state is shown in Fig.~3, which appears to be uniformly distributed.
For the case of horizontal polarisation, the phase distributions of both the vacuum state and a single photon are no longer uniform.
If one ignores (by tracing out) the polarisation dependence, for our examples the phase distributions become equal to the circular polarised ones in both figures.

Let us reconsider some properties of the polarisation-sensitive phase distributions from the viewpoint of Dirac's conjecture.
For the circular polarisation, the annihilation operator $\hat a_{+}$ is given by $\hat a_{+}\propto \hat q+{\rm i}\hat p$.
The anti-circular polarisation is obtained by replacing the momentum with the oppositely directed momentum $-\hat p$.
Thus the annihilation operator for the compound system of both polarisations can be written as
\begin{align}
	\hat {\bf a}=\hat a_{+}\otimes |\sigma_{+}\rangle\langle\sigma_{+}|+{\rm i}\hat a_{+}^\dagger\otimes |\sigma_{-}\rangle\langle\sigma_{-}|.
\nonumber
\end{align}
In fact, this yields formally the same expression as obtained by Dirac's proposal: $\hat {\bf a}=\hat {\bf V}\hat {\bf n}^{\frac{1}{2}}$, together with $\sqrt{-1}={\rm i}$.
Thus the eigenstates states for $\hat {\bf a}$, $\hat {\bf a}^\dagger$ of the two circular polarisations factorise with respect to the polarisation.

Based on the definition of the Hermitian phase operator together with its eigenvalue problem we are now able to define an observable Hermitian time operator as
\begin{align}
\hat {\bf T}=\frac{1}{\omega}\, \hat {\bf \Phi}.
\end{align}
It is directly related to the frequency $\omega$ of the monochromatic radiation field under study.
Obviously, it fulfils the commutation relation 
\begin{align}
[\hat {\bf T},\hat {\bf H}]={\rm i}\hbar\hat {\bf 1}.  \nonumber
\end{align}
Consequently we obtain that the often only vaguely discussed energy-time 
uncertainty relation,
\begin{align}
\Delta E\Delta T\geq\frac{\hbar}{2},
\nonumber
\end{align}
as a strict consequence of Heisenberg's uncertainty principle within the formalism of quantum physics.

In conclusion we have shown that a common description of the photon and its polarisation leads to a solution of Dirac's phase problem.
In a formal sense our approach resembles Dirac's quantum description of the relativistic electron.
In this manner a Hermitian phase operator can be defined that obeys the basic relations according to the conjecture by Dirac.
In the resulting phase probability distributions polarisation interference effects occur. In general, they are important for the phase statistics of a given quantum state.
As an important consequence, the developed approach to Dirac's quantum phase problem renders it possible to define a Hermitian operator for the observation of time, from which one rigorously obtains the energy-time uncertainty relation.

\noindent{\bf Acknowledgement}\\
Valuable discussions with T. Kiesel are greatfully acknowledged.
This work has been supported by the Deutsche Forschungsgemeinschaft through SFB 652.

\end{document}